\newcommand{\boldgamma}{\mbox{\boldmath $\gamma$}}

\newcommand{\boldsigma}{\mbox{\boldmath $\sigma$}}

\newcommand{\boldmu}{\mbox{\boldmath$\mu$}}

\newcommand{\eq}[1]{Eq.~(\ref{#1})}
\def\poinc{Poincar\'{e} }
\def\bfq {{\bf q}}\newcommand{\bb}{\langle}
\newcommand{\kk}{\rangle}

\def\bfK{{\bf K}}

\def\bfs{{\bf s}}\def\bfn{{\bf n}}
  \def\bfq {{\bf q}}\def\bfb{{\bf b}}
\def\bfr {{\bf r}}\def\bfR {{\bf R}}

\def\bfr{{\bf r}} \def\bfR{{\bf R}}

\def\be{\begin{equation}}
 \def \ee{\end{equation}}
\def\bea{\begin{eqnarray}}
  \def\eea{\end{eqnarray}}

\documentclass[
    ,final            
 ]
  {aipproc}

\layoutstyle{8x11single}

\begin{document}

\title{Overview of nucleon structure }



\author{Gerald A. Miller} {address={Physics Department, University of
Washington, Seattle, Washington 98195
}
}

\begin{abstract}
The quark-gluon properties of the nucleon are probed by a host of 
recent and planned experiments. These involve 
elastic, deep-inelastic, semi-inclusive deep-inelastic (SIDIS), 
and deeply-virtual Compton scattering. A light-front description
 is naturally applied to interpret all of these processes. The 
advantages of this approach will be discussed The talk will then
 focus on the transverse charge and magnetization densities
 of the nucleons and the use of SIDIS to reveal the non-spherical
 shape of the nucleon. We find that the central charge density of
 the neutron is negative, that the magnetization density of the
 proton extends further in space than the charge distribution, 
and that  the shape of the nucleon, as measured by the spin-dependent
density,
 is not spherical.
\end{abstract}

\maketitle


\section{Introduction}
The purpose of this talk is to illustrate the new  ways that
physicists are using to examine nucleon structure. New information is
originating from recent measurements of electromagnetic form factors, and
new
relations between these form factors and deep inelastic scattering DIS
encoded through generalized parton distributions
 GPDs, and transverse momentum distributions. I've chosen a few  
 examples, based on my experience, to represent some of the  
recent progress related to these quantities.

\medskip
The outline of the remainder of this presentation is as follows.
I'll begin with a brief introduction and discussion of the motivation
for studying the nucleon. Then 
I'll discuss some phenomenology \cite{Frank:1995pv,Miller:2003sa}
indicates that the proton is not round. Phenomenological methods will
be replaced by model-independent tools related to the transverse
charge density
to discuss the 
    neutron charge density, 
     proton magnetization  and the 
     shape of proton (which can also be observed experimentally 
\cite{Miller:2007ae}). It should be stated explicitly that this kind of
     theoretical work is driven by the recent vast experimental
     progress at various electron-scattering facilities around the world.
    
\medskip    
The main motivation for studying nucleon structure is simply to
     understand confinement. One addresses questions such as:
 How does the nucleon stick together
when struck by photon? 
 Where is charge and magnetization 
density located? 
What is the 
origin of the proton angular momentum? 
What is the shape of the proton?

\section{Lepton-nucleon scattering}
Electron and muon nucleon scattering has been used to probe the
nucleon for a very long time.
 Hofstatder \cite{Hofstadter:1956qs} used elastic scattering to
 discover
 that the nucleon was not not an elementary particle. Later
 inelastic scattering experiments at high momentum transfer (deep
 inelastic scattering) discovered
 the quarks within the nucleon \cite{ Bloom:1969kc}. We shall use
 information from  both kinds of experiments in this talk.

\medskip
There are two form factors that enter in elastic scattering because
the nucleon carries both charge and magnetization density. The
photon-nucleon vertex function $\Gamma_\mu$ is written as
\bea \Gamma_\mu=\gamma_\mu F_1(Q^2)+{i\sigma_{\mu\nu}\over 2M}F_2(Q^2).\eea
The Dirac $F_1$ and Pauli $F_2$ form factors are the fundamental
objects.
Sachs \cite{Sachs:1962zzc} introduced the linear combinations
\bea&&
G_E(Q^2)\equiv F_1(Q^2)-{Q^2\over 4M^2}F_2(Q^2)\\
&&G_M(Q^2)\equiv F_1(Q^2)+F_2(Q^2)\eea
to simplify the expression for the cross-section, obtained in
one-photon exchange approsximation. The form factors are our first
example.

\section{Elastic Form Factors and Their Interpretation}
The first point to realize is that, despite textbook statements,
 the electric form factor
$G_E(Q^2)$ is not the Fourier transform of the charge density 
\cite{Miller:2009sg,Miller:2009fc}.
The separation between relative and center of mass coordinates that is
 used in non-relativistic physics is not valid if the constituents
 move relativistically. In this case, the initial and final states
 have different total momentum and therefore have different wave
 functions. The square of a wave function does not appear in the form
 factor, and therefore  no charge densities appear.  Technically, we
 say that the effects of the boost between the rest frame and
 different moving frames must be included.
   
\medskip
A relativistic treatment is needed.
The most widely used and most effective approach is the use of 
 light front coordinates,  which is very closely related to working in
 the infinite     momentum frame. In a frame moving with velocity
 nearly that of light along the negative $z$ direction the usual time
 variable is changed to  
\bea
x^+=(ct+z)/\sqrt{2}=(x^0+x^3)/\sqrt{2},\eea
and its evolution
operator is given by \bea p^-=(p^0 +p^3) /\sqrt{2}.\eea In the
infinite momentum frame
$x^+$ plays the role of time and $p^-$ plays the role of the Hamiltonian.
The longitudinal 
spatial coordinate is given by 
\bea x^-=(x^0 -x^3)/\sqrt{2},\eea
 and the canonical momentum is given by
\bea p^+=(p^0+p^3)/\sqrt{2}.\eea
Transverse position, and momentum, {\bf b} , {\bf p} are the usual $x$
and
$y$ components. 
These light-front spatial and momentum 
coordinates are used to analyze form factors, and deep inelastic 
scattering.  GPDs, and TMDs are defined in terms of these variables.

\medskip
Only one other aspect of relativistic formalism is needed to proceed:
the  
kinematic subgroup of the \poinc group.   
 Lorentz transformations  in the transverse direction are
 kinematic. With a Lorentz trasnformation defined by a transverse 
velocity ${\bf v}$ , the variable $k^+$ is  unchanged, but 
$ {\bf k}\rightarrow {\bf k} -k^+{\bf v}$.
This is just like the non-relativistic Galilean  transformation except
that  $k^+$
takes the role of a  
mass. This means ultimately that if one 
takes the  momentum transfer to be in the perpendicular  
direction, the density is a two-dimensional 
Fourier transform. The condition on the momentum transfer is simply
expressed as \bea
q^+=(q^0+q^3)/\sqrt{2}=0,\label{qplus}\eea so that 
$-q^2 =Q^2={\bf q}^2.$ This condition can always be met for space-like photons.

\medskip
The relation \eq{qplus} is essential in trying to interpret the 
 form factor as a measurement of the charge density. To have a density,
 one needs  the 
 overlap of wave functions
 involving Fock space components with the same number of partons. The 
overlap of wave function components with different number of 
constituents does not have a probability or charge density
 interpretation. 
These are absent in a frame with $q^+=0$ 
because conservation of the plus-component of momentum 
prevents the creation of quark-pairs.

\section{Phenomenology}
I begin by discussing the 
model calculation of \cite{Frank:1995pv}. This work used a 
model  proton wave function expressed in terms of light-front
\poinc-invariant  variables. 
The use of  light front variables, and the conditon \eq{qplus} 
 incorporate the effects of the boost. This wave function is like the 
 usual quark model. But Dirac spinors, which carry orbital angular
 momentum
 replace the Pauli spinors. This model predicted the rapid fall, with
 increasing values of $Q^2$, of the
 ratio
$G_E/G_M$ of proton form factors \cite{Perdrisat:2006hj},
 and the related flat nature of the
 ratio 
$QF_2/F_1$ Ref.~\cite{Miller:2002qb}. This is shown in Fig.~1.
 The basic model feature
 that led to this result is the 
 lower components of the Dirac spinors.

\begin{figure}\label{fig6}
\unitlength1cm
  \includegraphics[angle=270,scale=0.3]{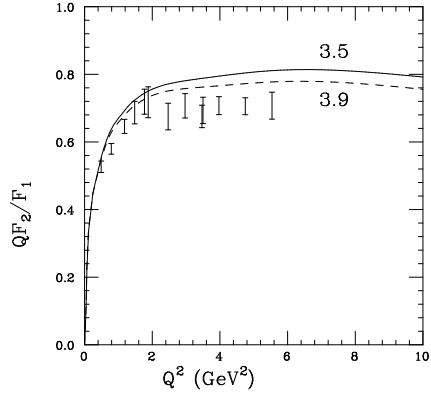}
\caption{The ratio $QF_2/F_1$, using two 
different values of a model parameter. 
The curves from the 1995 theory of \cite{Frank:1995pv}
for the ratio  are labeled by the
value of a  model parameter.
The  data 
are from 
\cite{Jones:1999rz} and 
\cite{Gayou:2001qd}. Figure reprinted with permission from 
\cite{Miller:2002qb}.}
\label{ratio}
\end{figure}
\vskip0.25cm\noindent 

\medskip
Thus we had a model which incorporated
quark   orbital angular momentum. This in turn led to the question:
does the proton have a non-spherical shape? The
 Wigner Eckart  theorem tells us that the spin 1/2 proton can have no
 quadrupole moment. However   one can define  
spin dependent densities SDD that do reveal a non-spherical shape
\cite{Miller:2003sa,Kvinikhidze:2006ty}.  The trick, when using
non-relativistic quantum mechanics,  is to replace the
usual usual density operator $\sum_i\delta(\vec{r}-\vec{r}_i)$ by the
spin dependent density
 $\sum_i1/2(1+\vec{\sigma}\cdot\bfn)\delta(\vec{r}-\vec{r}_i)$
 which gives the probability that a spin-1/2 constituent
 located at a position $\vec{r}$ has a spin in an arbitrary direction 
denoted by the unit vector 
$\bfn$. Please see Fig.~2 of \cite{Miller:2003sa} for a display of
the 
possible shapes of the nucleon, which include sphere, sausage,
peanut and doughnut. This is  also presented 
 in a popular publication \cite{Miller:2008sq}.

\medskip
It is worthwhile to present a simple example that illustrates 
 the connection between the  
 spin-dependent density and 
orbital angular momentum,
We consider the case of  a
 single charged particle
moving in a fixed, rotationally-invariant potential in an energy eigenstate
$|\Psi_{1,1,1/2,s}\rangle$ 
of quantum
numbers: $l=1,j=1/2$, polarized in the  direction $\widehat{\bfs}$
 and radial wave function $R(r_p)$. The wave function can be written
as 
\bea 
(\bfr_p|
 \Psi_{1,1,1/2,s}\rangle=R(r_p)\boldsigma\cdot\hat{\bfr}_p|s\rangle.
\label{wave}
\eea
The ordinary density. 
$\rho(r)=\langle\Psi_{1,1,1/2,s}|\delta(\bfr-\bfr_p)|
\Psi_{1,1,1/2,s}\rangle=R^2(r)$, is  spherically symmetric  because 
$(\boldsigma\cdot\hat{\bfr})^2=1$.
But 
the matrix element of 
the SDD \bea\rho(\bfr,\hat{n})=\langle \Psi_{1,1,1/2,s} 
\left\vert \widehat{\rho}(\bfr,\hat{n})\right\vert \Psi_{1,1,1/2,s} 
\rangle\eea
is more interesting:
\bea \rho(\bfr,\hat{n}) 
={R^2(r)\over 2}\bb  
\widehat{s}\vert\boldsigma\cdot \hat{\bfr}(1+\boldsigma\cdot 
\hat{n})\boldsigma\cdot\hat{\bfr}
\vert \widehat{s}\kk.\eea
\medskip
Previous publications chose the direction of the axis to be the $z$
direction, but we now are aware that transverse densities  are
measurable. This is because model-independent results can be obtained
from an infinite momentum frame analysis in which the $z$ direction is
that of the infinite momentum.
Therefore we consider the example in 
which the polarization is in the x-direction:
$|\hat{s}=\hat{x}\rangle=
{1\over\sqrt{2}}(|\uparrow\rangle+|\downarrow\rangle)$,
and the general unit vector $\hat{n}$ is replaced by $\bfn$  
which points   transversely to the
$z$ direction. Evaluation of the matrix element  for positions in the  
transverse $z=0$ plane gives 
shows that 
\bea \rho(\bfr,\bfn) =R^2(b^2){1\over
2}\left(1+\cos(2\phi-\phi_n)\right),\eea
where 
$b^2=x^2+y^2$, $\phi$ is the angle between $\bfr$ and the $x$-axis,
and $\phi_n$ is the angle between $\bfn$ and the $x$-axis.
Thus the transverse spin dependent density has an unusual cloverleaf shape.
as shown in Fig.~\ref{densityplot}.
 Please see below for brief
 discussions about how shapes related to these can  be measured and
 computed using lattice QCD techniques.

\begin{figure}\label{densityplot}
  \includegraphics[height=.3\textheight]{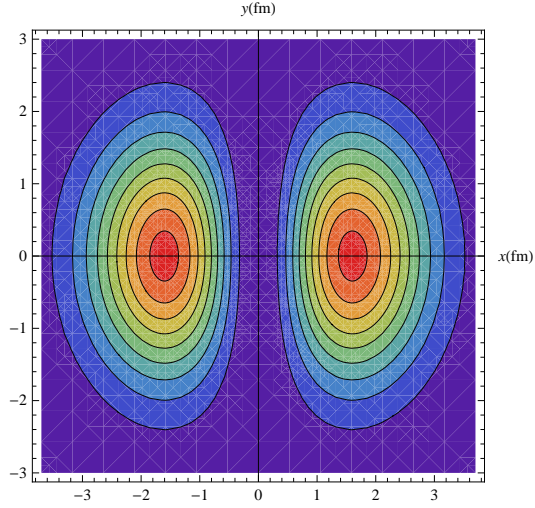}
  \caption{[Color Online] Spin-dependent 
transverse density of the model in the text. The lines are
  constant transverse density profiles, with the brighest areas
  indicating the maxima. }
\end{figure}
\section{Model independent transverse charge density}
In the infinite momentum frame the charge density operator becomes the 
plus-component of the current operator:
\begin{eqnarray}
\widehat{\rho}_\infty(x^-,{\bf b})=\Sigma_q e_q\bar{q}(x^-,{\bf b})\gamma^+
{q}(x^-,{\bf b})=J^+(x^-,{\bf b}).
\end{eqnarray}
The light-cone ``time'', $x^+=0$. The separation $\bfb$ can be defined
only if an origin can be established. This is done by
fixing  the position of the transverse center-of-momentum to be ${\bf
0}$, accomplished through the superposition of trasnverse momentum states:
\begin{eqnarray}
\vert p^+,{\bf R=0},\lambda\rangle={ N}\int {d^2p\over (2\pi)^2}\vert p^+,{\bf p},\lambda\rangle.
\end{eqnarray}
The density itself is defined as the  diagonal matrix element:
\begin{eqnarray}
\rho_\infty(x^-,{\bf b})=\langle  p^+,{\bf R=0},\lambda\vert\widehat{\rho}_\infty(x^-,{\bf b})\vert
p^+,{\bf R=0},\lambda\rangle
\end{eqnarray}
Using $J^+(x)=e^{i{\widehat P}\cdot x }J^+(0)e^{-i{\widehat P}\cdot x }$
 and the relation 
$F_1(Q^2)=\langle p^+,{\bf p}',\lambda\vert J^+(0)\vert p^+,{\bf p},\lambda\rangle$ leads to the main result
\begin{eqnarray}
\rho(b)\equiv \int dx^-\rho_\infty(x^-,{\bf b})=\int {d^2q\over
 (2\pi)^2}F_1(Q^2={\bf q}^2)e^{-i{\bf q}\cdot{\bf b}}. 
\end{eqnarray}
This is the key result of the talk. The transverse charge density
$\rho(b)$ is the two-dimensional Fourier transform of the Dirac form
factor, $F_1$.

\medskip
We begin the analysis by posing a 
motivational question. What is charge density at the center of the
neutron?
The neutron has no charge, but its charge density need not vanish.
Is the central density positive or negative? According to an old idea
presented by Fermi, a neutron sometimes   fluctuates into a proton
and a $\pi^-$. The proton, 
 as a heavy particle stays near the center, and the  $\pi^-$ 
floats
 to the edge. This was evaluated more recently using the cloudy bag
 model \cite{Thomas:1981vc}. Another idea is that the repulsive
 one-gluon-exchange 
 interaction between 2$d$ quarks favors configurations with a
 positively charged $u$ quark at the center 
\cite{Friar72,Carlitz:1977bd,Isgur:1980hh}. Both models suggest that
the central region of the neutron is positively charged and both 
provide a
function
$G^n_E(Q^2)$ that  rises a the value of 
$Q^2$ increases from 0.    

\begin{figure}\label{prl1}
  \includegraphics[height=.3\textheight]{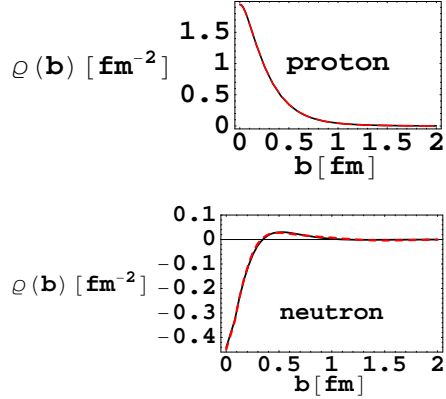}
  \caption{(Color online) Upper panel: proton charge density $\rho(b)$.Lower panel: neutron charge density. The solid curves use the  parameterization of
\cite{Kelly:2004hm}, and the dashed (red) curve uses \cite{Bradford:2006yz}.}
\end{figure}

\begin{figure}\label{prl2}
  \includegraphics[height=.3\textheight]{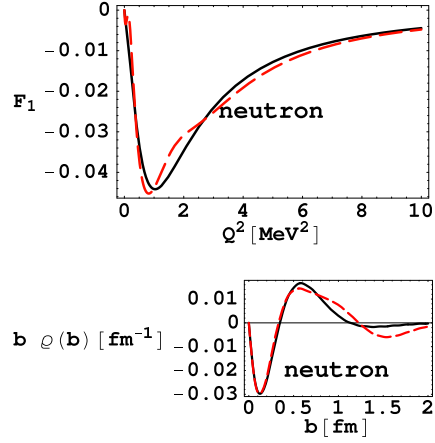}
  \caption{Upper panel: $F_1$. Lower panel:   $b \rho(b)$ in transverse position space. The
solid curves are  obtained using \cite{Kelly:2004hm} and the dashed curves with  \cite{Bradford:2006yz}. }
\end{figure}

\medskip 
  Here we avoid
the use of models by analyzing the transverse charge density, $\rho(b)$.
See figures 1,2 of \cite{Miller:2007uy}, which are shown here as
Fig.~\ref{prl1} and Fig.~\ref{prl2}.
Fig.~\ref{prl1} shows the surprising feature that the central charge
density of the neutron is negative.

\medskip
That the  central charge density of the neutron 
is negative  can be seen immediately from the negative-definite
nature of the measured values of $F_1(Q^2)$. 
The detailed structure of the neutron transverse charge density is
also interesting, as shown in Fig.~\ref{prl2}
The neutron transverse
charge density is negative at the center, positive in the middle and
negative at the outer edge. This suggests the presence of the pion
cloud.

\subsection{Explanation of the neutron's negative central charge
density}
The negative nature of the central neutron charge density cries out
for explanantion. This was attempted in different ways 
\cite{Miller:2008jc,Rinehimer:2009yv,Rinehimer:2009sz}    
Ref.~\cite{Miller:2008jc} used generalized parton distributions,
GPDs, to provide an interpretation.
 These are capable of yielding a three-dimensional picture of the
nucleon in which one determines probability densities for a quark to
be at a position $\bfb$ and carry a given value of the
longitudinal momentum transfer 
       \cite{Burkardt:2002hr}. For any given Fock-space component the
       relation
\bea \bfR={\bf 0}=\sum_i x_i \bfb_i,\eea 
where the sum over $i$ is over the  constituents in
the component. This tells us that  a quark carrying $x_i=1$ must be at
a postion $\bfb=0$.

\medskip
Various parameterizations 
\cite{Diehl:2004cx,Guidal:2004nd,Ahmad:2006gn,Tiburzi:2004mh} appear
in the literature. The common theme is that deep inelastic scattering
structure function $\nu W_2$
and the elastic form factors follow            the Drell-Yan
\cite{Drell:1969km}
\& West \cite{West:1970av}
relation:
\bea \lim_{Q^2\rightarrow\infty}F_1(Q^2)\propto{1\over
Q^{2n}},\;\lim_{x\rightarrow1}\nu W_2(x)=(1-x)^{2n-1}.\eea
The integer $n=2$ is one less than the number of valence quarks for a
nucleon and the relation is that the  same value of $n$ appears in the
high momentum transfer form factor and the $x$ near unity behavior of
the deep inelastic structure function. Thus large 
 values of $Q^2$ are related to
large values of $x$, which in turn means small values of $b$. This
connection between commuting position and momentum operators is not a
conseqeunce of the uncertainty principle.

\medskip   
The negatively-charged
 $d$ quark dominates the large $x$  distribution of the neutron,
\cite{Pumplin:2002vw} and this
which corresponds to small values of $b$. Therefore it is natural to
 conclude that the central negative charge density arises from
 $d$ quarks at the center of the neutron \cite{Miller:2008jc}.

\medskip
It is possible that these $d$ quarks arise from negatively charged
pions that penetrate the  interior of the nucleon. A simple
model in which the neutron fluctuates into a proton and a $\pi^-$,
parametrized to  reproduce the negative-definite nature of the
neutron's
$F_1$ \cite{Rinehimer:2009sz} reproduces the negative transverse central
density.  The change from the nominal, naive positive value of the charge
density 
obtained from $G_E$ can be understood as originating in the boost to
the infinite momentum frame \cite{Rinehimer:2009yv}.

    \section{Proton Magnetization density} 
The question of the location of the anomalous magnetization density of
the proton was examined in \cite{Miller:2007kt}. The magnetization
density was constructed by computing the matrix element of
$\boldmu\cdot{\bf B}$ in a nucleon polarized in the $x$-direction.
The result obtained is that the anomalous magnetization density
$\rho_M(b)$
is given by 
\bea \rho_M(b)=\int{d^2q\over (2\pi)^2}F_2(Q^2=\bfq^2)e^{-i\bfq\cdot\bfb}.\eea
If one parameterizes the low $Q^2$ form factors as 
\bea F_1(Q^2)\approx(1-Q^2\langle b^2\rangle_{CH}/4),\quad
F_2(Q^2)\approx(1-Q^2\langle b^2\rangle_{M}/4),\eea 
Ref.~\cite{Miller:2007kt} found that the mean square magnetization
 transverse radius $\langle b^2\rangle_{M}$ is greater than the 
mean square transverse charge radius $\langle b^2\rangle_{Ch}$:
\bea 
    \langle b^2\rangle_{M}-\langle b^2\rangle_{Ch}\approx {\mu\over
    M^2}>0.\eea
The question of the actual magnetization density was examined in
\cite{Miller:2010xx} with the result that the actual magnetization
    density is $-b_y{\partial \over \partial b_y}\rho_M(b)$. In either
    case the proton magnetization density extends further than its
    charge density.

      \section{Measuring the shape of the proton}
    The experiments capable of measuring the proton shapes reported
    above
 are studied in \cite{Miller:2008sq}.
The field-theoretic spin-dependent density is related to the
    transverse momentum
distribution
 TMD, 
      $h_{1T}^\perp$.
TMDs are defined generally through the relations \cite{Mulders:1995dh}
\bea \Phi^{\Gamma}(x,\bfK)=
\int {d\xi^-d^2\boldmath{\xi}\over2(2\pi)^3} e^{iK\cdot\xi}
\langle P,S|\bar{q}(0)\Gamma{\cal L}q(\xi,\xi^+=0)|P,S\rangle,
\eea
where ${\cal L}$ is an appropriate gauge link-operator
and specifically  we use
\bea  \Phi^{i\sigma^{i+}\gamma_5}(x,\bfK)=S_T^i h_1(x,\bfK^2)+
{(K_T^iK_T^j-{1\over2}\bfK^2\delta_{ij})S_T^j\over
M^2}h_{1T}^\perp(x,\bfK)
\eea 
as the light-front version of the spin-dependent density.
Here $x\equiv K^+/P^+$. The relation with an equal-time density
operator is achieved by integrating the above quantities over
$x$. This sets $\xi^-=0$ so that $x^\pm=0,z=0,t=0$.

The quantity $h_{1T}^\perp$ can be measured using polarized electrons
in the reaction $\vec{e}p\rightarrow e'\pi X$. Other reactions can be
used also; see the references in \cite{Miller:2008sq}. If $\int dx
h_{1T}^\perp\ne0$, the proton is not round.

\section{Computing the shape of the proton using lattice QCD}
One can define general model-independent densities \cite{Diehl:2005jf}
by considering
matrix elements of the form $\bar{q}(0,\bfb)\gamma^+ \Gamma q(0,\bfb)$,
where $\Gamma$ is a Dirac matrix. The choice of $\Gamma$ corresponding
to the spin-dependent density discussed earlier is
 $\Gamma={1\over2}(1+\bfs\cdot\boldgamma\gamma^5),$  where $\bfs$ is an
 arbitrary vector representing the direction of the quark spin.
 Evaluating the matrix element of this quantity
produces the spin-dependent density, which can be thought of as 
the $x^-$ integrated version of the coordinate space results of
 \cite{Miller:2003sa}.
The proton is non-spherical if the function $\tilde{A}''_{T10}$ of
\cite{Gockeler:2006zu} is non-vanishing, as recent lattice
calculations show \cite{Schierholz:2009xx}. Thus the proton is not round.

      \section{Summary}
We have tried to explain how 
elastic and deep inelastic electron scattering 
can be analyzed from a unified light=front frame work using transverse
charge densities, generalized parton distribution and transverse
momentum
distributions.
Key physics results are that the neutron central charge density is
negative, the proton magnetization density extends further than its
charge density and that future experiments
can determine whether or not the proton is round, as predicted by
models \cite{Miller:2003sa} and QCD \cite{Schierholz:2009xx}
 by extracting the
quantity
$h_{1T}^\perp$.




\begin{theacknowledgments}
 I thank the USDOE for partial support of this work, and my
 collaborators for all their copious help.
\end{theacknowledgments}



\bibliographystyle{aipproc}   

\bibliography{millertalk}

\IfFileExists{millertalk.bbl}{}
 {\typeout{}
  \typeout{******************************************}
  \typeout{** Please run "bibtex \jobname" to optain}
  \typeout{** the bibliography and then re-run LaTeX}
  \typeout{** twice to fix the references!}
  \typeout{******************************************}
  \typeout{}
 }

\end{document}